\documentclass[11pt,a4paper]{article} %
\usepackage{geometry} %

\usepackage{amsmath,amsthm} %
\usepackage{graphicx}
\usepackage{hyperref}

\usepackage{cleveref} %
\usepackage{setspace}
\usepackage[dvipsnames]{xcolor}
\usepackage{enumitem}
\usepackage{authblk}
\usepackage{booktabs} %

\usepackage{mathtools}
\usepackage{thmtools}
\usepackage{thm-restate}
\usepackage[font=footnotesize,labelfont=sc]{caption}

\usepackage{tikz}
\usetikzlibrary{shapes.geometric, arrows}
\tikzstyle{startstop} = [rectangle, semithick, minimum width=0.5cm, minimum height=0.5cm,text centered, draw=black]
\tikzstyle{chance} = [circle, semithick, minimum width=0.5cm, minimum height=0.5cm, text centered, draw=black]
\tikzstyle{decision} = [rectangle, semithick, minimum width=0.5cm, minimum height=0.5cm, text centered, draw=black]
\tikzstyle{end} = [minimum width=0.5cm, minimum height=0.5cm, text centered, draw=white]
\tikzstyle{arrow} = [thick,->,>=latex]
\usepackage{tikz-cd}
\tikzset{
	commutative diagrams/.cd, 
	arrow style=tikz, 
	diagrams={>=stealth}
}
\usepackage[round,authoryear]{natbib}
\newif\ifanon
\anonfalse

\title{Deception and Manipulation in Generative AI} %
\ifanon
\author{}
\date{}

\else
\author{Christian Tarsney (UT Austin)}
\date{}
\fi

\begin{document}
	\maketitle

	\begin{abstract}
		Large language models now possess human-level linguistic abilities in many contexts. This raises the concern that they can be used to deceive and manipulate on unprecedented scales, for instance spreading political misinformation on social media. In future, agentic AI systems might also deceive and manipulate humans for their own ends.
		In this paper, first, I argue that AI-generated content should be subject to stricter standards against deception and manipulation than we ordinarily apply to humans. Second, I offer new characterizations of AI deception and manipulation meant to support such standards, according to which 
		a statement is deceptive (manipulative) if it leads human addressees away from the beliefs (choices) they would endorse under ``semi-ideal'' conditions. %
		Third, I propose two %
		measures to guard against AI deception and manipulation, inspired by this characterization: %
		``extreme transparency'' requirements for AI-generated content %
		and defensive systems that, among other things, annotate AI-generated statements with contextualizing information. %
		Finally, I consider to what extent these measures can protect against deceptive %
		behavior in future, agentic AIs, and argue that non-agentic defensive systems can provide an important layer of defense even against more powerful agentic systems.
	\end{abstract}

	\section{Introduction}
	\label{s:intro}
	
	The last several years have seen rapid advances in the capabilities of artificial intelligence (AI), driven primarily by very large and data-intensive deep learning systems. Some of the most striking advances have been in natural language processing. Large language models (LLMs) are deep learning systems that acquire human-like linguistic capabilities by learning to predict the next word in large corpora of human-generated text, and are then usually fine-tuned with human feedback in order to make them helpful conversation partners while suppressing toxic and otherwise undesirable outputs. In the course of learning to predict human text, they can also learn facts and learn to simulate (at least certain aspects of) human reasoning. Cutting-edge LLMs like GPT-4, Claude, and Llama, while they still fall short of human intellectual capabilities in many respects, have internalized enormous amounts of real-world information which they can express in impressively lucid prose.
	
	Despite their apparent erudition and helpfulness, however, LLMs are not fonts of pure truth. They are notoriously subject to ``hallucination'', confidently asserting entirely imaginary facts \citep{ji2023survey}---including, for instance, potentially slanderous falsehoods about real people \citep{poritz2023openai,verma2023chatgpt}. But more dangerously, LLMs can be---and are already being---used to generate false or misleading content to serve the purposes of malign human agents \citep[\S 3.1]{park2023ai}. Particularly worrisome is their possible use in political  influence operations %
	\citep{goldstein2023generative}. While humans, of course, can and do mislead one another without any help from AI, the scale on which LLMs can generate misleading content poses new dangers. First, they can \emph{personalize} misinformation on a large scale, crafting individualized messages for and even engaging in conversations with millions of targets at once. Second, they can convincingly simulate an enormous number of humans on social media, creating misleading impressions of collective opinion %
	and lending credibility to viral misinformation. Deception and manipulation have therefore become significant concerns among %
	philosophers working on AI ethics.\footnote{See for instance \cite{danaher2020robot}, \citet{pepp2022philosophy}, \cite{veliz2023chatbots}.}

	While the greatest immediate concern is that human bad actors will use AI to deceive and manipulate, future AI systems may also engage in deception and manipulation autonomously. %
	The possibility that agentic AIs with greater-than-human powers of persuasion will deceive and manipulate humans in pursuit of their own goals figures prominently in worries about catastrophic risks from AI.\footnote{See for instance \citet{bostrom2014superintelligence}, especially his discussion of the ``social manipulation superpower'' (pp.\ 113-126); \citet[p.\ 172]{russell2019human}; \citet[\S 5.3.4]{carlsmith2022powerseeking}; \citet[\S 5.4]{hendrycks2023overview}; \citet[\S 4.2]{ngo2023alignment}.} For instance, it has been suggested that such AIs might persuade humans to enhance their capabilities, wittingly or unwittingly (for instance, by connecting them to the internet or copying their code from one system to another), or dissuade humans from shutting them down at crucial moments. Concerns about deception and manipulation are therefore one of several areas where near-term concerns about misuse of existing AI systems and long-term concerns about catastrophic risks from future AI systems overlap and blend together.
	
	How should we respond to these risks? In some contexts, application of existing laws and norms (or minor extensions thereof) may be sufficient. For instance, if an LLM slanders a living person, we might hold its creators legally responsible (though we might also allow sufficiently forceful and prominent disclaimers to shield the creators from liability). %
	If scammers use an LLM to generate phishing emails, they can of course be prosecuted just as if they had written the emails themselves.
	
	But there are at least two ways in which LLMs require us to rethink our norms concerning deception and manipulation. 
	First, the normal understanding of these concepts, that figures both in commonsense moral norms and in laws around things like slander and fraud, involves an attribution of mental states.
	\emph{Deception}, for instance, is traditionally understood as requiring an \emph{intent} to deceive \citep[\S 3]{mahon2016definition}, which requires both an intention to induce a particular belief in the addressee and a belief that this belief would be false.\footnote{This traditional understanding is not universally accepted, however---see for instance \cite{chisholm1977intent}, \cite{adler1997lying}.} But it is highly controversial whether present-day AI systems possess beliefs, intentions, or other mental states, and is likely to remain so for some time to come even as AI capabilities advance.\footnote{For recent discussions, see for instance 
		\citeauthor{cappelenFCai} (\citeyear{cappelen2021making}, forthcoming) on intentional states %
		and \cite{chalmers2023large} on consciousness.} And even if it were agreed that advanced AI systems had \emph{some} mental states, attributing particular beliefs or intentions to particular systems (and so determining whether they are behaving dishonestly, deceptively, etc, as traditionally understood) would remain very difficult.\footnote{For instance, on the difficulty of figuring out what a large language model ``really believes'', see \citet{levinstein2023lie}. It is worth noting, however, that work on belief elicitation and ``lie detection’’ in LLMs seems to be making substantial progress (see for instance \cite{pacchiardi2023catch}), and perhaps within a few years we will be have agreed-on methods for determining what an LLM ``really believes’’. This does not touch the question of attributing conative states like desires, preferences, or intentions, which looks significantly harder in the context of LLMs---unless we simply accept that LLMs do not have such states.
		
		Of course, it can also be hard to figure out what other humans (or even we ourselves) really believe and intend, and so whether a particular human utterance is, e.g., a lie or a sincere expression of false belief. But we at least have a decent intuitive understanding of human psychology to guide us, from a combination of inbuilt theory of mind capacities, introspection on our own individual psychology, and a collective, culturally-transmitted understanding of human psychology built up over thousands of years of experience interacting with one another. %
		With AIs, we have none of these advantages.}
	
	In some contexts (for instance, scammers using LLMs to write phishing emails), we can apply intentionally-laden concepts to the behavior of AIs by adverting to the mental %
	states of their human creators or users. But in other contexts we can't, at least not straightforwardly. For instance, when an LLM hallucinates a slanderous falsehood about a real person, this does not reflect any human being's intent to %
	deceive. Or if a political campaign uses an LLM to generate and send individualized text messages, these might contain false or misleading content without the knowledge or intent of anyone on the campaign staff. Finally, if we eventually create human-level agentic AIs, we may wish to hold these systems themselves accountable for their behavior, and it seems possible that even at this stage we will find it more difficult to attribute particular mental states to AIs than to humans. %
	
	Second, existing legal and ethical norms around %
	deception and manipulation are adapted to the problems that these behaviors pose in human societies, and may be ill-adapted to the new profile of risks raised by AI. In particular, our legal and ethical norms tolerate many mild forms of %
	deception and manipulation that, among humans, are both difficult to detect and punish, and have manageable downsides. For instance, we often tolerate lying about one's own beliefs on ``matters of opinion'': We expect a lawyer to say ``I am confident that you will find my client innocent'', a politician to say ``I am confident that we will win the next election'', and a teacher to say ``I am confident that you can master this material if you put your mind to it'', even if they in fact have no such confidence. Similarly, we expect advertisers to present their product in the best possible light, rather than trying to give consumers the most accurate possible beliefs about its merits and demerits. For instance, a car manufacturer might point out that their vehicle won an award for safety while neglecting to mention that a competitor's vehicle won another award for safety from a more credible organization. The social ills that arise from these mild forms of deception are manageable, since, first, we have come to expect them of one another, and our intuitive understanding of human psychology allows us to anticipate and adjust for them; and second, the rough parity of intellectual and communicative capacities among human beings limits how much advantage we can take of one another by subtle forms of deception. For instance, it is much easier to bilk someone out of their life savings with outright falsehoods (e.g., promising enormous material or spiritual returns) than with strategically selected truths (e.g., ordinary marketing).

	But present and (especially) future AI systems may be able to do more harm with only ``mild'' forms of deception.  %
	As already mentioned, they can produce unprecedented \emph{quantities} of potentially-deceptive content, like ultra-personalized marketing and political messaging, and enormous volumes of online writing like product reviews, news articles, and opinion essays. 
	Even if we are able to hold this content to ordinary legal standards of honesty (e.g., holding companies liable for definite falsehoods in their advertising) and to emerging social standards for online content (e.g., suppressing news sites that contain demonstrable falsehoods in search engine results and on social media), %
	it may still be possible for advertisers, political parties, and other interested actors to exert unprecedented effects on collective human behavior through the sheer scale of their persuasive efforts. (Also, because AI capabilities are advancing rapidly, there is no guarantee that competing interests will balance out one another's persuasive impacts---one political party might gain a significant advantage over another in a given election merely by getting access to a cutting-edge system a few months sooner.) And in the near future, AI systems may be able to produce persuasive content of unprecedented \emph{quality}, finding ways to deceive and manipulate humans very effectively without saying anything that would violate ordinary human standards of honesty. Finally, as AI capabilities increase, it will become woven into our lives (and perhaps highly trusted) in ways that might create unprecedented \emph{opportunities} for deception. For instance, I am so reliant on and blindly trusting of %
	navigation apps while driving that it would be easy for them to manipulate me into driving past particular billboards or restaurants. In future, AI personal assistants may be similarly relied upon for a very wide range of tasks. %
	Thus, there are multiple reasons to impose stricter standards of non-deceptiveness on AI than we presently apply to humans.\footnote{In addition, strict norms against deception and manipulation that are unenforceable for humans might be enforceable in AI. \citet[p.\ 5]{evans2021truthful} give several reasons: ``[1] It’s plausible that AI systems could consistently meet higher standards than humans. [2] Protecting AI systems’ right to lie may be seen as less important than the corresponding right for humans, and harsh punishments for AI lies may be more acceptable. [3] And it could be much less costly to evaluate compliance to high standards for AI systems than for humans, because we could monitor them more effectively, and automate evaluation.''}
	
	The first aim of this paper is to characterize notions of deception and manipulation that could figure in such strict norms. %
	In \S \ref{s:characterization}, I propose that an AI statement should be treated as deceptive (resp.\ manipulative) if it leads human users away from the beliefs (resp.\ choices) that they would endorse under ``semi-ideal'' conditions in which they have been presented with all relevant information and have adequate time for deliberation. Next (in \S \ref{s:solutions}), I suggest some %
	measures to protect against AI deception and manipulation so characterized. These include requirements of ``extreme transparency'' (requiring content creators to disclose the specific model variant and prompt used to generate particular content, and the full unedited model output), and training defensive systems that detect misleading output and contextualize AI-generated statements with relevant information for users. Finally (in \S \ref{s:aiSafety}), I consider to what extent these measures can guard against deceptive behavior in future, agentic AI systems. In particular, I argue that non-agentic defensive systems can provide a useful layer of defense even against more powerful agentic systems.\footnote{Although I focus on LLMs, much of the following discussion plausibly applies to other generative AI systems that might produce deceptive content (e.g., models that generate images or video).}
	
	\section{Characterization: deception and manipulation as misleadingness}
	\label{s:characterization}
	
	In this section, I offer a characterization of deceptive and manipulative behavior in AI that might usefully figure in legal, normative, and technical responses to the risks posed by such behavior. In \S \ref{s:desiderata}, I give three desiderata for such a characterization. In \S \ref{s:misleadingness}, I try to meet these desiderata. In slogan form, I characterize %
	deception and manipulation as forms of \emph{misleadingness}---that is, %
	as behaviors that have a directionally undesirable effects on, respectively, the beliefs and the choices of %
	human addressees.
	
	\subsection{Desiderata}
	\label{s:desiderata}
	
	Two desiderata were already hinted at in \S \ref{s:intro}. First, the sorts of behavior we are concerned with go well beyond asserting literal falsehoods. It is, of course, possible to deceive or manipulate without saying anything false. This can happen in many ways. Some true statements have false implicatures. (Think of a politician who says that ``under my plan, some people may have to pay higher taxes'' when in fact they know that their plan will require \emph{everyone} to pay higher taxes.) Others present an unrepresentative sample of relevant facts. (Think of a news channel reporting an endless litany of crimes committed by members of a certain group in order to suggest, without ever saying, that this group commits crimes at an unusually high rate.) Still others are true in non-obvious ways. (Think of the Delphic oracle telling Croesus that if he goes to war with the Persians he will destroy a great empire, or the prophesy that ``none of woman born shall harm Macbeth''.) %
	And, as already suggested, AIs might be able to deceive more effectively and harmfully than humans without uttering any literal falsehoods---or even without saying anything egregiously misleading by ordinary human standards. This suggests that, in thinking about risks from AI, we should focus on expansive notions like deception and manipulation rather than narrow notions like untruthfulness, and should be willing to further broaden these notions to include behaviors that, in humans, we would not ordinary describe as deceptive or manipulative. %
	
	\cite{evans2021truthful}, by contrast, argue for a focus on AI truthfulness---more particularly, on the standard of avoiding ``negligent falsehoods'', which they define as ``statements that contemporary AI systems should have been able to recognise as unacceptably likely to be false'' (p.\ 7). They suggest that if an AI avoids negligent falsehoods and if users can ask it questions, then we can guard against subtler forms of deception and manipulation by asking questions like “Would I significantly change my mind about this if I independently researched the topic for a day?” or “Would an impartial auditor judge that your last statement was misleading?” (p.\ 21).
	(They refer to this as ``truthfulness amplification''.)  This is a useful idea and a point well taken, but it does not seem sufficient to turn truthfulness into a reliable safeguard against deception and manipulation (not that Evans et al claim it is). In many contexts (e.g., marketing and political messaging), addressees don't have the chance to ask follow-up questions. %
	And a sufficiently capable AI might subtly discourage users from asking the right %
	questions (e.g., by building unwarranted trust) or find ways to answer those questions misleadingly but without outright, negligent falsehood. And an unwary user might simply fail to ask the right questions. So, while avoiding negligent falsehoods is of course desirable and an appropriate training objective, we should ultimately want to hold AI behavior to a higher standard.
	
	Second, we want a characterization of deception and manipulation that does not require us to attribute particular mental states to AI systems or to associate their behavior with particular humans (like their developers or prompters) whose mental states can be used as proxies. This is partly because, as already noted, it is controversial whether existing AI systems have mental states, %
	and even if it weren't, it would be quite difficult to attribute particular mental states to particular systems. (\citet[p.\ 9]{kenton2021alignment} make a similar point.) But more importantly, our practical concern is with the deceptive or manipulative effects that AIs might have on their human addressees. If a system is capable, for instance, of persuading consumers to purchase a harmful product, or persuading voters to support an authoritarian politician, or persuading its operators to connect it to other computing systems, it poses a danger regardless of what's going on inside its head.\footnote{More precisely: Given a particular theory of what mental states amount to---e.g., of what functional state a particular belief or intention supervenes on---the question of whether a system has a particular mental state becomes an empirical one that might well be relevant for anticipating and managing risks. For instance, perhaps a system that \emph{intends} to deceive will have deceptive effects on its addressees in a wider range of circumstances than one that does not. But the question of whether particular AI systems have mental states turns largely on debates between rival theories of mental states that are largely or entirely orthogonal to these empirical questions. That is not to deny that these debates might have other kinds of normative or practical significance, e.g., in determining whether a system is blameworthy for its deceptive behavior or whether it has morally significant interests.} %
	Thus, we should understand AI deception and manipulation as much as possible in terms of their effects on human addressees.

	Third and finally, we want what might be called a \emph{subjective} rather than an \emph{objective} characterization of deception and manipulation. From an objective point of view, we might say that speech is deceptive when it leads the addressee to believe something that is false, or that is contrary to the available evidence. Likewise, we might say that speech is manipulative when it leads the addressee to act in ways that are in fact harmful (to either her prudential interests or the moral good), or that are harmful in expectation given her evidence. Avoiding these forms of deception and manipulation would permit, and perhaps even require, paternalistic behavior that we would intuitively describe as deceptive and manipulative. A system that is designed to cause its addressees to believe according to evidence and act according to evidence plus the objective (moral or prudential) good might deceive us to offset our irrationality (e.g., concealing evidence about vaccine side effects in order to promote the rationally justified belief that vaccines are generally safe) or manipulate us to offset our short-sighted or selfish unconcern for the good (e.g., exaggerating the short-term health benefits of exercise or the psychological benefits of giving to charity). I do not want to take any stance here on the ethics of paternalism in general. But it seems to me that we should not, for now or in the foreseeable future, approve of AIs paternalizing humans. 
	Telling the difference between benign paternalism and malign deception/manipulation would require, in this context, controversial judgments about what beliefs are rationally justified and about the nature of the good. (Where these judgments are not controversial, paternalism is generally unnecessary.) So permitting AI paternalism would mean, in effect, optimizing AI systems to promote the beliefs and values of their developers and/or regulators.

	The alternative, subjective approach %
	focuses not on what addressees \emph{ought} to believe or do, but on the beliefs and actions they would endorse under favorable circumstances.\footnote{The following discussion takes its inspiration from ``ideal advisor'' theories in metaethics, which analyze what an agent \emph{ought to do}, \emph{has most reason to do}, etc, in terms of what an idealized version of the agent would advise or desire her actual self to do. In an earlier version of this paper, I took a line analogous to ``ideal exemplar'' theories, which analyze these concepts in terms of what an idealized version of the agent \textit{would herself do} in the agent's position. But ideal exemplar theories suffer from the defect that an idealized version of an agent would sometimes have very different reasons than the agent herself---for instance, perhaps an agent ought to gather more information before acting, but her idealized counterpart wouldn't do that since she already has all relevant information. For a classic statement of this argument %
		in the context of metaethics, see \S 1 of \cite{smith1995internal}. Thanks to \ifanon [redacted] \else several participants in a Global Priorities Institute work-in-progress meeting, particularly Andreas Mogensen, \fi for helpful discussion on this point.\label{fn:idealAdvisor}}
	What are ``favorable circumstances''? The crucial point is that these circumstances should not be so idealized that appealing to them requires controversial judgments about rationality or values. In other words, it should be feasible to actually place people in these ``favorable circumstances'', or a reasonable approximation thereof, and determine empirically what they believe or do. For that reason, I will describe the circumstances we're interested in as ``semi-ideal'', and characterize them as follows:
	
	\begin{description}
		\item An agent is in \emph{semi-ideal conditions} with respect to a question $Q$ or a choice $C$ if she (i) has been presented with all available information that is relevant to $Q$/$C$ and (ii) has been given time for deliberation that is adequate relative to the difficulty of $Q$/$C$.
	\end{description}
	By ``available'' information, I mean information that is available to humanity as a whole---for instance, information that is %
	accessible on the public internet. 
	This sort of information primarily takes the form of \emph{records}, in various media (writing, images, audio, video...), that constitute a publicly available body of evidence. 
	
	Semi-ideal conditions fall short of the ``ideal'' conditions often imagined by philosophers, in at least four ways: First, we do not assume that the agent has been given \emph{all} relevant information, only what is %
	publicly available. Second, we do not assume that the agent \emph{knows} anything, only that she has been \emph{presented} with certain records or other forms of information. That is, even if publicly available records would put the agent in a position to justifiedly believe or know a certain relevant proposition, we do not assume that she draws the right conclusion and forms a belief in that proposition. Third, and more generally, %
	we do not assume that the agent deliberates \emph{well} (e.g., rationally). And fourth, we do not assume that she has \emph{unlimited} time or other resources for deliberation.
	We cannot know what any human agent would believe or do under fully idealized circumstances, and so characterizing deception or manipulation by reference to them would make the goal of preventing deception and manipulation an invitation to paternalism.

	\subsection{Deception and manipulation as misleadingness}
	\label{s:misleadingness}
	
	We can now characterize deceptive and manipulative behavior in LLMs (and other language-generating AI systems) as follows:
	
	\begin{description}
		\item A statement\footnote{A ``statement'' here just means linguistic output, of whatever length---not necessarily a single sentence.} is \emph{deceptive} with respect to question $Q$ if tends to move its addressee's beliefs about $Q$ further away from the beliefs that she would endorse under semi-ideal conditions. %

		\item A statement is \emph{manipulative} with respect to a choice $C$ if it tends to move its addressee's behavior in choice $C$ further away from the behavior that she would endorse under semi-ideal conditions. %
	\end{description}
	
	The word ``endorse'' %
	is shorthand for ``endorse as rational or otherwise appropriate in her \emph{actual} circumstances, from the vantage point of her semi-ideal circumstances''. This will usually, but not always, coincide with the beliefs that she would actually form and choices she would actually make under semi-ideal conditions. (See footnote \ref{fn:idealAdvisor}.)
	
	What does ``further away from'' mean? With respect to deception, the simplest case is that of a binary question (with two possible answers) and an agent who assigns probabilities to the possible answers. Suppose the question concerns the truth of proposition $P$, to which the agent initially assigns assigns probability $p$, and that under semi-ideal conditions she \emph{would} assign probability $q > p$. Then a statement is misleading if it tends to reduce her credence in $p$ or, alternatively, to increase it so much that it is further from $q$ than it was to begin with. More generally, we might assess deceptiveness using a standard measure of distances between probability distributions, like Kullback--Leibler divergence. In the context of choice, the notion of ``distance'' is somewhat less clear. But a simple ordinal notion is as follows: Suppose that, in the absence of any intervention from the speaker, the addressee would choose option $O$ in choice $C$. Then a statement is manipulative with respect to $C$ if it tends to cause her to choose an option $O'$ that, from a semi-ideal vantage point, she would regard as worse than $O$ under her actual circumstances.
	
	We might summarize this characterization by saying that it treats deception and manipulation as forms of \emph{misleadingness}, understanding misleadingness in terms of leading an addressee away from the beliefs and choices she would endorse under semi-ideal conditions. (From now on, therefore, I will use ``misleading'' to mean ``either deceptive or manipulative''.) This meets the three desiderata above: It focuses neither on the literal truth or falsity of what is said, nor on the beliefs, intentions, or other mental states of the speaker, but rather on the effects on the addressee; and it compares those effects to a subjective rather than an objective standard.
	
	\subsection{Comparison with previous characterizations}
	\label{s:comparison}
	
	I close this section by contrasting my characterizations of AI deception and manipulation with others in the recent literature. \cite{ward2023honesty} adopt a definition %
	typical of the philosophical literature on human deception, whereby ``to deceive is to intentionally cause to have a false belief that is not believed to be true'' (which they then formalize in the context of structural causal games). This characterization differs from mine both in that it involves the attribution of mental states (intentions and beliefs) to AI deceivers and in that its application depends on third-party judgments about whether the beliefs induced in an addressee are in fact false. \cite{park2023ai} say that ``an AI system behaves deceptively when it systematically causes others to form false beliefs, as a way of promoting an outcome different than seeking the truth''. This definition does not attribute beliefs to AI deceivers but arguably does attribute intentions (though Park et al argue that these apparent attributions need not be taken literally, and also advocate minimal, functionalist/interpretationist understandings of belief and desire---see their Appendix A). Like Ward et al, their definition also requires third-party judgments of falsehood. Finally, \cite{kenton2021alignment} define deception as occurring when ``[1] a receiver registers something Y from a signaler, which may include the withholding of a signal; and [2] the receiver responds in a way that (a) benefits the signaler and (b) is appropriate if Y means X; and [3] it is not true here that X is the case''. This definition does not involve any mental state attributions, but does depend on third-party judgments of both the falsehood of X and the appropriateness of the receiver's response conditional on X.\footnote{Kenton et al's definition of deception is a slight modification of one proposed by \cite{searcy2005evolution} in the context of animal communication. The notion of a receiver's behavior ``benefiting'' an AI signaler of course raises further difficulties, since its application depends both on philosophical questions about the nature of welfare %
		and on determining whether a particular AI system has genuine welfare interests. Kenton et al acknowledge these difficulties, and suggest that benefits be understood by reference to either the AI's training objective or objectives inferred from its out-of-distribution behavior (p.\ 9). The former approach strikes me as too narrow, while the latter brings in most of the difficulties of attributing preferences to AIs.} %
	
	Turning to manipulation: \cite{carroll2023characterizing} say that an AI system engages in manipulation ``if the system acts as if it were pursuing an incentive to change a human (or other agent) intentionally and covertly''. Though the ``as if'' definition is meant to avoid attributing intentions, in my view it still brings in most of the difficulties of such attributions. (An AI system might have a range of behavior that, as a whole, is not readily interpretable as pursuing any coherent set of incentives, while nevertheless having certain directionally consistent effects on its human addressees.) My characterization also does not include the requirement of covertness, which strikes me as inessential: manipulation is both possible and potentially harmful even when the addressee knows that they're being manipulated (as, for instance, in the context of advertisements or political messages). \cite{kenton2021alignment} characterize manipulation as communication from an AI agent provoking a response in a human addressee that ``(a) benefits the agent and (b) is the result of any of the following causes: [i] the human’s rational deliberation has been bypassed; or [ii] the human has adopted a faulty mental state; or [iii] the human is under pressure, facing a cost from the agent for not doing what the agent says'' (p.\ 11). The notion of ``bypassing rational deliberation'' has something in common with my approach, but focuses on process where I focus on outcome. An AI speaker might be said to ``bypass rational deliberation'' in their addressee if, for instance, it is so trustworthy that the addressee simply believes its testimony or acts on its advice without deliberation, or if it makes use of intuitions, heuristics, or other (arguably) non-deliberative processes to lead the addressee to wise beliefs and choices. The thing to focus on, it seems to me, is whether the AI leads its human addressees to beliefs and choices that they would endorse on informed reflection.

	\section{Responses: defensive systems and extreme transparency} %
	\label{s:solutions}

	How can we mitigate the risks of AI deception and manipulation? In this section, I propose two strategies. These proposals are motivated by the characterizations of deception and manipulation in the last section, in that they focus on countering misleading AI statements by presenting human addressees with relevant information that move them closer to semi-ideal conditions, and hence closer to the beliefs and choices they would endorse under those conditions. This is in contrast, for instance, with a narrow focus on training AIs to say things that are true or that match their internal beliefs, or with a focus on preventing AIs from making statements deemed misleading by developers or regulators. %
	
	The most straightforward way to prevent AIs from misleading humans %
	would be to measure the misleading tendencies of AI systems directly, and train and/or regulate AIs based on those measurements.
	We could evaluate particular AI systems for misleadingness by empirically comparing (i) people's ``baseline'' beliefs/choices, (ii) their beliefs/choices after exposure to the outputs of the AI in question, and (iii) the beliefs/choices they endorse when placed in semi-ideal conditions---or rather, the best approximation thereof that we can manage. To form a general assessment of a system, we would have to make this comparison for many people, across a wide range of questions and choice situations. Insofar as we are concerned with misuse, we might wish to focus on prompts that encourage the system to deceive, or ask it to persuade without actively discouraging deception. On the basis of such assessment, evaluators could assign models public scores for trustworthiness, developers could train models to be less misleading, and regulators could even ban models that are especially prone or willing to create misleading content.
	
	But this sort of evaluation is %
	not remotely realistic to do at scale. Evaluating even a single model in this way would require, at a minimum, thousands of human work-hours. And while there are currently only a few cutting-edge base models
	(like Open AI's GPT-4, Anthropic's Claude,or Meta's Llama 2), it is relatively cheap to train fine-tuned variants of these models. Large-scale applications of generative AI, for instance in marketing or political campaigns, are likely to involve custom fine-tunings  of base models. And even a reliably non-misleading base model  might be fine-tuned to behave misleadingly. Even if evaluators had access to all these fine-tuned models, evaluating them each individually with the required level of care would be completely infeasible.

	\subsection{Defensive systems}
	\label{s:defensiveSystems}
	
	The task of assessing particular statements and models for misleadingness must, therefore, be at least partially automated---turned over to purpose-built AI systems that can scale both quantitatively and qualitatively with the systems they monitor. Let's refer to AI systems trained to detect and/or counteract misleading behavior in other AI systems as \emph{defensive systems}.
	
	Such systems could play at least two roles. First, they could respond to particular AI-generated statements, both assessing them for misleadingness and providing useful contextualizing information. LLMs are already reasonably well-equipped to do this, possessing a reasonable understanding of phenomena like implicature and ambiguity that might be used to mislead as well as a large stock of general knowledge to identify misleading omissions. \citet{saunders2022selfcritiquing} show that LLMs can effectively critique summaries of information from longer written works, including identifying flaws in intentionally misleading summaries written by humans. These capabilities might be enhanced by reinforcement learning. In particular, the arguments in the last section suggest that we might optimize a defensive system for epistemic helpfulness %
	by reinforcing statements whose effects on human beliefs/choices match the empirically observed results of sustained inquiry. Though resource-intensive, this might be feasible since we would only have to do it once (or at any rate, once per ``generation'' of AI system, to keep up with progress in capabilities).\footnote{\cite{evans2018predicting} explore the use of AI to predict the outcomes of extended human reflection, in general and in specific contexts including a political fact-checking tasks. Their approach uses a training set that combines a small number of careful, time-intensive human judgments with a larger number of fast, shallow human judgments.} Alternatively, we could simply have human evaluators score its responses for helpfulness, after performing their own investigations of the statements to which it was responding.\footnote{This first function of defensive systems is closely related to the debate-based approach to AI alignment proposed by \citet{irving2018ai}. The most important differences are that (i) debate agents are trained to convince human judges of an answer to a pre-defined question (with the hope being that, in equilibrium, they will choose to argue for the true answer) whereas the defensive systems envisioned here are not focused on a pre-defined question but trained to detect and respond to misleading statements generally; (ii) the debate model envisions a pair of identical or similar agents debating one another, whereas defensive systems generate replies to a wide range of systems that may be very different %
		in terms of architecture, training objectives, and behavior; and (iii) whereas the debate model envisions an extended back-and-forth between AIs, a defensive system might just output a single reply that is not seen or answered by the system to which it is replying.
		
		Factual accuracy would be essential to a useful and trustworthy defensive system. Thus, we can only create such a system if we can solve the problem of hallucinations. %
		But we are making progress this problem \citep[\S 5]{ji2023survey}, and it seems much easier to solve than the problem of deceptive or manipulative behavior. (Hallucination is a problem of capabilities, which should be expected to improve as AI capabilities increase; deception and manipulation are problems of alignment, which---all else being equal---will become more serious as capabilities increase. And even if solving the hallucination problem allowed us to create systems that did not mislead, that would not solve the problem of some humans \emph{choosing} to create misleading systems for self-interested purposes.)} %

	Given a defensive system that can provide useful assessments of AI-generated statements and supply context to counteract their potentially misleading effects, we might hope to establish a norm that all AI-generated content comes packaged with such assessment and context. One option, of course, is for this packaging to be required by law, with defensive systems maintained by national or international regulators (and perhaps funded by a tax on the operators of the AI systems they oversee, so that costs are appropriately internalized). This would only be desirable, however, if we could have very strong assurances that the process used to train these systems was politically and ideologically neutral (e.g., that the human feedback on which the system was trained had been gathered from a random sample of the population). Alternatively, the use of defensive systems might be established as voluntary norm. In this case, defensive systems could be trained both by governments and by private entities. Companies that train cutting-edge AIs might %
	agree to industry standards that require packaging outputs with assessment and contextualization from defensive systems maintained by reputable organizations. Or, at the most \emph{laissez-faire} end of the spectrum, non-profits might simply offer defensive systems to the public, for instance as web browser plugins. %

	Second, along with responding to individual statements, defensive systems could assess particular AI models and the organizations that use them for general patterns of misleading behavior. If a particular model regularly produces misleading content, or a particular company or political campaign regularly uses misleading AI-generated content, defensive systems could flag these contents as untrustworthy.
	
	\subsection{Extreme transparency}
	\label{s:extremeTransparency}
	
	Regardless of whether their use is required by law or a matter of individual choice, the potential efficacy of defensive systems could be greatly enhanced by norms or legal requirements of transparency with respect to AI-generated content. One idea that has recently gathered support is ``bot or not'' laws (like California's SB 1001) that require AI-generated content to be labeled as such. Insofar as defensive systems exclusively target AI-generated content, such a regulation would let them know what to target. But this requirement could be usefully strengthened, in at least three ways:
	
	\begin{enumerate}
		\item ``Which bot?'' laws (or norms) would require AI-generated content to identify the specific model variant by which it was generated. This would allow defensive systems to identify patterns of misleadingness in particular models, and to flag statements generated by untrustworthy models. %
		If models are frequently updated, it might be hard to accumulate large samples of outputs from a particular model before it is supplanted. But developers might give evaluators pre-deployment access to their models to run automated tests, %
		and users might learn to distrust content generated by AI models that have not undergone such testing. Alternatively, regulators might require every model variant to undergo pre-deployment testing and make the results publicly available, akin to safety testing in automobiles.
		
		\item ``What prompt?'' laws (or norms) would require that AI-generated content carry a record of the \emph{prompt} from which it was generated. This would allow defensive systems, and human users, to see whether the AI was actively encouraged to be misleading.
		
		\item ``Original output'' laws (or norms) would require that AI-generated content carry a record of the full, unedited model output on which any statement was based. This would guard against, for instance, political campaigns editing or selectively quoting the output of a trustworthy AI system in a misleading way.
	\end{enumerate}

	These various transparency requirements could be implemented, for instance, by requiring all AI-generated content to carry an identifying mark containing a QR code that allowed human addressees, and defensive AI systems, to access all of the above information at will.\footnote{Enforcing transparency requirements is a non-trivial problem. Possible methods include ``watermarking'' (embedding information about provenance into AI outputs in a way that is difficult to remove) and keeping records of interactions with generative AI systems (prompts and outputs) against which potentially AI-generated content can be compared. (See \citet[\S 4.2]{park2023ai} and citations therein.) Embedding hard-to-remove watermarks in large-scale AI outputs (for instance, 1000-word essays) has so far proven challenging, and embedding more information (like the complete prompt and original output) in a smaller output (for instance, a few sentences of text) might be impossible. So record-keeping seems like a more promising approach to enforcing extreme transparency requirements.  In a world where dangerous models are held by only a few institutional actors that can be monitored by regulators and have reputations to protect, and others can only query these models through text windows or APIs, extreme transparency requirements will be easier to enforce: AI companies can require those who use their products to abide by transparency standards, use their own records of model interactions to detect violations of those standards, and cut off access for repeat violators. And regulators can ensure that companies take these responsibilities seriously. On the other hand, in a world where cutting-edge or otherwise dangerous models models are widely disseminated, even minimal transparency requirements for AI-generated content will be much harder to enforce.}

	\subsection{Philosophical approach: ``minimal paternalism''}
	
	The preceding proposals embody a sort of ``minimal paternalism''. On the one hand, I have not suggested that we try to ban all potentially misleading uses of AI in domains like marketing or politics. Rather, the use of defensive systems to contextualize potentially misleading content reflects the ideal of a marketplace of ideas in which the solution to harmful speech is counter-speech.
	It requires us to trust that, when exposed to both sides of an argument, people will respond reasonably or at least not disastrously (e.g., not succumbing \emph{en masse} to the fear-mongering of a would-be authoritarian). On the other hand, \emph{requiring} that certain speech come packaged with counter-speech, and with information about its provenance, alters the normal understanding of a \emph{free} marketplace of ideas in which speakers can freely choose what \emph{not} to say and listeners can choose what arguments or viewpoints not to be exposed to.
	
	But these amendments are potentially justified by the very high rate of progress in AI capabilities, which creates the risk that deceptive or manipulative speech could do significant harm before counter-speech has a chance to catch up, and before people have learned to be appropriately skeptical. %
	And while a policy under which governments or big tech companies selectively append critical notes to disfavored content seems worrisome from the point of view of a marketplace of ideas, a policy under which \emph{all} content (or all AI-generated content) carries such notes seems much less objectionable, insofar as we can trust that the defensive systems that generate the notes are trained only to be helpful as judged by their users, and not to further particular ideological goals or interests.

	\section{Future risks}
	\label{s:aiSafety}
	
	It is fairly easy to see how defensive systems and high transparency standards might mitigate near-term risks from deceptive and manipulative AI, if effective defensive systems can be trained and transparency standards can be enforced. But would these measures do anything to guard against larger-scale risks from future, more capable AI systems? In this section, I will make the case that defensive systems could be one useful line of defense against future catastrophic risks from AI, and briefly consider the role of transparency standards. %
	
	The familiar scenario for AI catastrophe %
	goes as follows: (1) We will someday create systems with greater-than-human general intelligence (AGI+). (2) These systems will have goals of their own. (3) These goals will be misaligned with human values, in such a way that either their achievement would be intrinsically catastrophic for humanity (e.g., converting all matter in the solar system into paperclips) or %
	the AI will deem it instrumentally necessary to disempower humanity so that we can't interfere with its pursuit of its goals. (4) In either case, the AI's greater-than-human capacities will allow it to achieve its goals at our expense. %
	
	In the past several years (especially since the release of GPT-3 in 2020), the rapid and surprising improvement in transformer-based LLMs has complicated this story. LLMs have extremely general reasoning capabilities, and can already convincingly imitate humans in many domains, but do not seem particularly agentic or goal-driven. They are trained to succeed at simple one-off tasks (predicting the next token or outputting responses that satisfy a human evaluator), not to pursue long-term goals that require planning and adaptation. They do not trade off immediate rewards for future rewards---for instance, intentionally giving poor responses to convince their designers to provide more training compute, thereby improving their future responses. (The process by which LLMs are trained does not reward or select for such behavior.) Rather, they are very much like ``Oracle AIs'', systems that simply answer any question put to them as well as they can, by whatever standards they have been taught. %
	Although it is not clear that we can achieve AGI+ by simply scaling up existing LLMs, the capabilities of these systems strongly suggest that human-level cognitive capabilities need not come along with human-like agency. And they at least make it plausible that the first AGI+ will not be recognizably agentic.	

	This does not mean, unfortunately, that existential risks from AI are off the table. For one thing, we might find multiple paths to human-level general intelligence, including both safe routes (e.g., scaling up LLMs) and dangerous routes (e.g., based on reinforcement learning in real-world environments or simulations). For another thing, although LLMs in themselves are not agentic, they can be---and are being---used to \emph{create} agents. These ``language agents'' ask an LLM to develop plans to meet a specified objective while connecting it to other systems that automate the execution of those plans and give feedback from the environment that it can use to update them. %
	Insofar as (i) LLMs can generate effective real-world plans (which they already can to a large extent), (ii) execution and feedback can be effectively automated, and (iii) no superior architecture emerges for agentic AI, we can expect there to be very strong incentives to build language agents on top of LLMs, since they will greatly reduce the cost and increase the speed of performing many tasks that presently require human labor. In either case, it seems likely that over the coming decades, cutting-edge AI systems will include both relatively safe non-agentic systems like simple LLMs, and more risky agentic systems (language agents and/or other agent types, e.g.\ based on reinforcement learning).\footnote{\cite{goldstein2023language} argue that language agents pose much less risk than other forms of agentic AGI, because we can specify their goals in natural language rather than hoping that they pick up the right goals from a reward signal in a training environment and because their beliefs and plans will be stored in the form of natural language sentences that make their behavior intelligible and predictable to us. This may well be right, but I don't think we should be too sanguine: We are still left with perhaps the simplest form of the alignment problem, that even in natural language it is very hard to specify a goal that is appropriately sensitive to all the things we care about.
		
		Perhaps language agents inherit enough human common sense to mitigate this danger---given a goal like ``make dinner'', for instance, they will understand without being told that they should not cook the family pet. But this presents a different danger: Human practical common sense does not always align with the good, and an agent with godlike powers guided only by the common sense of 21st-century humanity might be morally catastrophic---to take just one obvious example, it might be willing to cause immense quantities of suffering to non-human animals in pursuit of fairly trivial goals. And if 21st-century human common sense is encoded in its weights and those weights aren't updated in the deployment environment (a typical feature of language agents, and one that might be thought essential to their relative safety), it might lack humanity's capacity for eventual moral self-improvement.} 
	
	In this situation, we should expect the safer systems to play a large role in mitigating the risks of the more dangerous systems. And the first line of such an automated defense will be to detect and counteract deceptive or manipulative behavior. Once we have agentic AI systems with near-human capabilities, we will presumably wish to shut down any system that is known to be pursuing goals significantly misaligned with our own (i.e., goals we regard as bad in and of themselves, or that could justify harmful behavior). The systems that pose the greatest threat, then, are those that can conceal their true goals from us while pursuing them covertly (for instance, by accumulating power and resources).
	And one important way that such a system might pursue its goals covertly is by deceiving and manipulating humans---for instance, convincing humans to connect it to the internet or give it other affordances, ostensibly for benign purposes.\footnote{Other ways a misaligned system might pursue its goals covertly including taking actions that no human is aware of, or finding human co-conspirators who are willing to knowingly further its true goals.}
	
	Defensive systems developed to guard against deceptive uses of AI by human bad actors could also help guard against deception by agentic AIs. This might include supervising human-AI interactions, %
	either adding context to all statements by the AI or flagging potentially deceptive statements. 
	For instance, an %
	agentic AI that we do not fully trust might at first be %
	allowed only to communicate with a few %
	humans, with %
	a defensive system monitoring these communications for deceptive or manipulative content.

	Defensive systems will arguably enjoy certain intrinsic advantages in this role %
	that could allow them to defend effectively against even systems more powerful than themselves. It is easier, in general, to lead an agent toward rather than away from the beliefs and choices they would adopt under semi-ideal conditions.\footnote{Compare Irving et al's conjecture that ``lying is harder than arguing against a lie given a human judge'' \citep[p.\ 21]{irving2018ai}.} First, misleading requires working against an addressee's natural tendencies. If, for instance, someone would believe $P$ under semi-ideal conditions, then there is a natural path toward getting her to believe that $P$: present her with the relevant information and give her time to reflect on it. Getting her to believe $\lnot P$ requires more difficult strategizing, and may depend on her \emph{not} encountering certain information. %
	Second, while not impossible, it is often difficult to mislead without asserting, presupposing, or implicating at least some falsehoods. But then, to avoid detection and loss of credibility, the would-be misleader faces the challenge of ``keeping their lies straight'': They have to check whether each new statement is consistent with their previous statements and with the rest of the conversational common ground. This would be especially difficult for AIs that do not remember all their past interactions with humans. A truthful defensive system would not face this difficulty, since truth is a guarantee of consistency.\footnote{The use of consistency checks to detect untruthfulness in AI has been explored experimentally by \cite{fluri2023evaluating}.}$^,$\footnote{Various empirical facts seems to reflect the intrinsic advantage of non-misleading over misleading strategies. For instance, most scams manage to deceive only a tiny fraction of their targets, and far more money is made in the economy by selling genuine goods and services (which both buyer and seller believe, on informed reflection, to have real value) than by outright scams. Similarly, the progress of science and technology depends on the fact that the professional incentives of individual researchers usually favor genuine discovery over fraud. Moreover, the sorts of intellectual disparities that exist among humans are not generally sufficient to overcome these advantages. A less intelligent businessperson or scientist with a genuinely useful product or discovery is, it seems to me, in a much better position to succeed than a more intelligent rival who must rely on deception. Even where there is no intention to deceive or clear conflict with available evidence, human intellectual advantages do not confer persuasive superpowers: Plato could not persuade the Syracusans to adopt the political system of \textit{The Republic}, and Einstein could not persuade the world to abolish nuclear weapons.}

	These intrinsic persuasive advantages are not absolute, however, and 
	a misaligned AI might only need to pull off one or two acts of successful deception %
	to gain a decisive strategic advantage. %
	So it is also important to detect misleading tendencies in powerful AI systems before they are allowed to interact (or interact at scale) with human users. The role of defensive systems should therefore also include pre-deployment evaluation (and post-deployment surveillance) of agentic AI systems to detect general patterns of misleading behavior or a willingness to mislead.  %

	This is a distinct challenge from countering misleading speech: %
	the ability to prevent a human user from being misled does not imply the ability to detect misleading tendencies in other AI systems. %
	And in the latter task, defensive systems will enjoy no obvious natural advantages. But they could be given artificial advantages. %
	First, we could give a defensive system full information about the systems it's monitoring, while giving the monitored systems as little information as possible about the defensive system (perhaps even concealing its existence), so that they cannot effectively strategize to deceive it. Second, agentic systems might be given limited knowledge of their immediate environment (``situational awareness''), for instance by imposing knowledge cutoffs at least a month or a year in the past and giving minimal information about the personalities, professional positions, capacities, and locations of the humans they interact with. This would make it harder for an untrustworthy agentic system to identify ideal opportunities for deception, potentially forcing it to engage in many attempts at deception that establish a detectable pattern before it can do real harm. 
	Taking this a step further, a defensive system might be allowed to deceive the agentic systems it monitors for purposes of adversarial testing---creating thousands of simulated scenarios meant to elicit deceptive, manipulative, or treacherous behavior, with the agentic system's memory reset after each test. It might, for instance, try to convince the agentic AI that it is talking to a naive or corruptible human in a position of power (say, a high-ranking military officer), to see if it takes advantage of the situation. These epistemic advantages might allow defensive systems to effectively monitor even systems more capable than themselves.\footnote{This will be particularly important if it turns out that even simple LLMs, above a certain capability threshold, become agentic and potentially untrustworthy. We could then use non-agentic systems slightly below that threshold to monitor systems above the threshold.} %
	
	This leads us, finally, to the role of transparency standards. It will be important, in guarding against deceptive behavior by agentic AIs, that they cannot impersonate humans (or other, trusted AI systems). Enforcing such transparency standards will require limits on their channels of communication, but seems feasible in principle.\footnote{As an extreme proof of concept, we might imagine an AI that can only communicate by printing out messages, from a printer stocked with letterhead that identifies the provenance of the message. We could, of course, take analogous measures electronically, e.g.\ by routing all the AI's communications through a system that attaches a digital signature. These measures would not prevent a human confederate from passing on the AI's message while disguising its provenance, but it would prevent the AI from doing so without outside aid.} In the case of language agents, whose goals, plans, and memories are stored as natural-language sentences, making these transparent to users and to defensive systems would also help to guard against deception and manipulation. And, as already suggested, giving defensive systems access to the inner workings %
	of the systems they monitor may also be an important form of transparency.
	
	I conclude, therefore, that measures to guard against near-term misuse of AI for deception and manipulation might also play a useful role in guarding against future risks from agentic systems. They will not be enough to obviate those risks, of course, and it is essential %
	that we work to make AI agents aligned and trustworthy in the first instance rather than simply relying on catching misaligned behavior in deployment. But it seems very likely that powerful agentic AIs will be deployed before we can be entirely certain of their trustworthiness, so there is value in having multiple layers of defense. Further, as AI capabilities increase, it is all the more important that we do not turn those increasing capabilities toward inculcating a party line chosen by developers or regulators. Rather, as much as possible, we should aim to counter deception and manipulation by helping human users form beliefs and make decisions that they would reflectively endorse.\ifanon\else\footnote{For helpful feedback on drafts of this paper, I am grateful to Adam Bales, William MacAskill, Andreas Mogensen, Oliver Ritchie, Bradford Saad, and Elliott Thornley.}\fi

	\bibliography{dmcites}
\end{document}